%% file: main.tex
\documentclass[sigconf]{acmart}

\usepackage{hyperref}
\usepackage{enumitem}
\usepackage{graphicx}
\usepackage{subcaption}

\newcommand{\PatchVec}{CC2Vec}

\copyrightyear{2020} 
\acmYear{2020} 
\setcopyright{acmcopyright}
\acmConference[ICSE '20]{42nd International Conference on Software Engineering}{May 23--29, 2020}{Seoul, Republic of Korea}
\acmBooktitle{42nd International Conference on Software Engineering (ICSE '20), May 23--29, 2020, Seoul, Republic of Korea}
\acmPrice{15.00}
\acmDOI{10.1145/3377811.3380361}
\acmISBN{978-1-4503-7121-6/20/05}

\begin{document}

\title{\PatchVec: Distributed Representations of Code Changes}


\author{Thong Hoang, Hong Jin Kang, David Lo}
\affiliation{%
  \institution{Singapore Management University, Singapore}
  }
\email{{vdthoang.2016, hjkang.2018, davidlo}@smu.edu.sg}

\author{Julia Lawall}
\affiliation{%
  \institution{Sorbonne University/Inria/LIP6, France}
  }
\email{Julia.Lawall@inria.fr}

\input{abstract}
\maketitle

\input{intro}

\input{approach}
\input{experiment}
\input{discussion}

\input{related_work}
\input{conclusion}

\input{ack}

\balance
\bibliographystyle{ACM-Reference-Format}
\bibliography{bib}
\end{document}

%% file: abstract.tex
\begin{abstract}
    Existing work on software patches often use features specific to a
    single task.  These works often rely on manually identified features, and
    human effort is required to identify these features for each task.  In
    this work, we propose CC2Vec, a neural network model that learns a
    representation of code changes guided by their accompanying log
    messages, which represent the semantic intent of the code changes.
    CC2Vec models the hierarchical structure of a code change with the help
    of the attention mechanism and uses multiple comparison functions to
    identify the differences between the removed and added code.

    To evaluate if CC2Vec can produce a distributed representation of code
    changes that is general and useful for multiple tasks on software
    patches, we use the vectors produced by CC2Vec 
    for three tasks: log message generation, bug fixing patch identification,
    and just-in-time defect prediction.  In all tasks, the models using CC2Vec
    outperform the state-of-the-art techniques.

\end{abstract}

%% file: intro.tex
\section{Introduction}
\label{sec:intro}

Patches, used to edit source code, are often created by developers to describe new features, fix bugs, or maintain existing functionality (e.g., API updates, refactoring, etc.). 
Patches contain two main pieces of information, a log message and a code change. 
The log message, used to describe the semantics of the code changes, is written in natural language by the developers.
The code change indicates the lines of code to remove or add across one or multiple files.
Research has shown that the study of historical patches can be employed
to solve software engineering problems, such as just-in-time defect
prediction~\cite{kamei2016studying, hoang2019deepjit}, identification of
bug fixing patches~\cite{tian2012identifying, hoang2019patchnet}, tangled
change prediction~\cite{kirinuki2014hey}, recommendation of a code reviewer
for a patch~\cite{rahman2016correct}, and many more.



Exploring patches to solve software engineering problems requires choosing a representation of the patch data. 
Most prior work involves manually crafting a set of features
to represent a patch and using these features for further
processing~\cite{tian2012identifying, kamei2016studying,
  mockus2000predicting, kamei2012large, kim2008classifying,
  yang2015deep}. These features have mostly been extracted from properties
of patches, such as the modifications to source code (e.g., number of removed and added
lines, the number of files modified), the history
of changes (e.g., the number of prior or recent changes to the updated
files), the record of patch authors and reviewers (e.g., the number of
developers or reviewers who contributed to the patch), etc. 
These features can be used as an input to a machine learning classifier (e.g., Support
Vector Machine, Logistic Regression, Random Forest, etc.) to address various
software engineering tasks~\cite{kamei2016studying,tian2012identifying, kirinuki2014hey,rahman2016correct}. 
Extracting a suitable vector representation to represent the ``meaning'' of a patch is certainly crucial. 
Intuitively, the quality of a patch representation plays a major role in determining the eventual learning outcome.



In this paper, to boost the effectiveness of existing solutions 
  that employ the properties of patches,
we wish to learn vector representations of the code changes in patches that can be used for a number of tasks. 
We propose a new deep learning architecture named \PatchVec~ that can effectively embed
a code change into a vector space where similar changes are close to each
other. 
As log messages, written by developers, are used
to describe the semantics of the code change, we use them to supervise the
learning of code changes' representations from patches. 
Specifically, \PatchVec~ optimizes the vector representation of a code
  change in a patch
  to predict appropriate words, extracted from the
first line of the log message. We consider only the first line, as
it is the focus of many prior works~\cite{liu2018neural,
  rahman2015textrank}, and is considered to carry the most semantic meaning
with the least noise.\footnote{https://chris.beams.io/posts/git-commit/}

\PatchVec~ analyzes the code change, i.e., scattered
fragments of removed and added code across multiple files. 
Code removed or added from a file follows a hierarchical structure (words form line, lines form hunks). 
Recent work has suggested that the attention mechanism can help in modelling structural dependencies~\cite{DBLP:conf/iclr/KimDHR17,alon2019code2vec}, 
thus, we hypothesize that the attention mechanism may be effective for modelling the structure of a code change.
We propose a specialized hierarchical attention network (HAN) to construct a vector representation of the removed code (and another for the added code) of each affected file in a given patch. 
Our HAN first builds vector representations of lines; these
vectors are then used to construct vector representations of hunks; 
and we then aggregate these vectors to construct the embedding vector of the removed or added code. 
Next, we employ multiple comparison functions to capture the 
difference between two embedding vectors representing removed and added code. 
This produces features representing the relationship between the removed and added code. 
Each comparison function produces a vector and these vectors are then concatenated to form an embedding vector for the affected file. 
Finally, the embedding vectors of all the affected files are
concatenated to build a vector representation of the code change in a patch. 
After training is completed, 
CC2Vec can be used to extract representations of code changes even from patches with empty or meaningless log messages (which are common in practice~\cite{jiang2017automatically, liu2018neural, linares2015changescribe}).
CC2Vec is also programming-language agnostic; one can use it to learn vector representations of code changes for any language.

The code change representation enables us to employ the power of (potentially a large number of) unlabeled patch data to improve the effectiveness of  
supervised learning tasks (also known as semi-supervised learning~\cite{chapelle2009semi}). 
We can use the code change representation to boost the effectiveness of many supervised learning tasks (e.g., identification of bug fixing patches, just-in-time defect prediction, etc.), especially on those tasks 
for which only a limited set of labeled data may be available. 

\PatchVec~ converts code changes into their distributed representations
by learning from a large collection of patches. The distributed representation captures pertinent features of the code changes by considering the characteristics of the whole collection of patches. 
Such distributed representations can be used as additional features for other tasks. 
Past studies have demonstrated the value of distributed representations to improve text classification~\cite{mikolov2013distributed}, action recognition~\cite{maji2011action}, image classification~\cite{cheung2012convolutional}, etc. 
Unfortunately, prior to our work, there is no existing solution that can produce a distributed representation of a code change. 

To evaluate the effectiveness of \PatchVec, we employ the representation
learned by \PatchVec~ in three software engineering tasks: 1) log message generation~\cite{liu2018neural} 2) bug fixing patch identification~\cite{hoang2019patchnet}
and 3) just-in-time defect prediction~\cite{hoang2019deepjit}. 
In the first task of log message generation, we generate the first line of a log message given a code change. 
\PatchVec~ can be used to improve over the best baseline by 24.73\% in terms of BLEU score (an accuracy measure that is widely used to evaluate machine translation systems).
For the task of identifying bug fixing patches, \PatchVec~ helps to improve the best performing baseline by 5.22\%, 9.18\%, 4.36\%, and 6.51\% in terms of accuracy, precision, F1, and Area Under the
Curve (AUC).  
For just-in-time defect prediction, \PatchVec~ helps to improve the AUC metric by 7.03\%
and 7.72\% on the QT and OPENSTACK datasets~\cite{mcintosh2017fix} as compared to the best baseline. 

The main contributions of this work are as follows: 
\begin{itemize}
	\item We propose a deep learning architecture, namely \PatchVec, that learns distributed representations of code changes guided by the semantic meaning contained in log messages. To the best of our knowledge, our work is the first work in this direction.
	
	\item We empirically investigate the value of integrating the code change vectors generated by \PatchVec~ and feature vectors used by state-of-the-art approaches on three tasks (i.e., log message generation, bug fixing patch identification, and just-in-time defect prediction) and demonstrate improvements.
\end{itemize}

The rest of this paper is organized as follows. 
Section~\ref{sec:approach} elaborates the design of \PatchVec. 
Section~\ref{sec:experiment} describes the experiments that demonstrate the value of our learned code change representations to aid in the three different tasks. 
Section~\ref{sec:discussion} presents an ablation study and some threats to validity. 
Section~\ref{sec:related_work} describes related prior studies. 
We conclude and mention future work in Section~\ref{sec:conclusion}.

%% file: approach.tex
\section{Approach}
\label{sec:approach}

\begin{figure}[t!]
	\centering 
	\includegraphics[width=0.775\columnwidth]{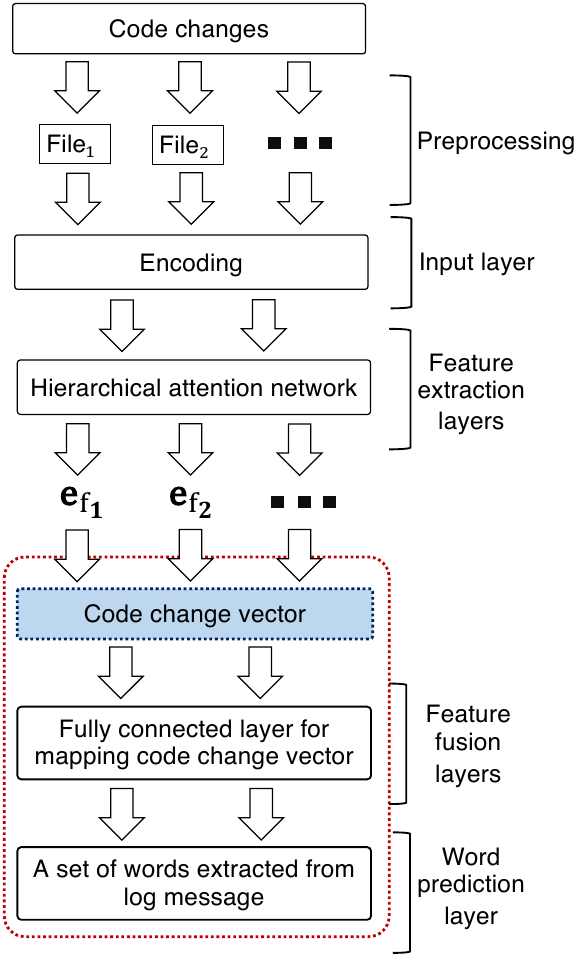}
	\caption{The overall framework of \PatchVec. Feature extraction layers are used to construct the embedding vectors for each affected file from a given patch (i.e., $\textbf{e}_{\texttt{f}_1}$, $\textbf{e}_{\texttt{f}_2}$, etc). The embedding vectors are then concatenated to build a vector representation for the code change in the patch (code change vector). The code change vector is connected to the fully connected layer and is learned by minimizing an objective function of the word prediction layer.}
	\label{fig:framework}
\end{figure}

In this section, we first present an overview of our framework. We then describe the details of each part of the framework. Finally, we present an algorithm for learning effective settings of our model's parameters.

\subsection{Framework Overview}
\label{sec:framework_overview}
 
Figure~\ref{fig:framework} illustrates the overall framework of \PatchVec. \PatchVec~ takes the code change of a patch as input and generates its distributed representation. 
\PatchVec~ uses the first line of the log message of the patch to supervise learning the code change representation. Specifically, the framework of \PatchVec~ includes five parts: 
\begin{itemize}		
	\item {\textit{Preprocessing}}: This part takes information from the code change of the given patch as an input and outputs a list of files. Each file includes a set of removed code lines and added code lines.
	\item {\textit{Input layer}}: This part encodes each changed file as a three-dimensional matrix to be given as input to the hierarchical attention network (HAN) for extracting features. 
	\item \textit{Feature extraction layers}: This part extracts the embedding vector (a.k.a. features) of each changed file. The resulting embedding vectors are then concatenated to form the vector representation of the code change in a given patch.
	\item \textit{Feature fusion layers and word prediction layer}: 
	This part maps the vector representation of the code change to a word
        vector extracted from the first line of log message; the word
        vector indicates the probabilities that various words
          describe the patch.
\end{itemize}

\PatchVec~ employs the first line of the log message of a patch to guide the
learning of a suitable vector that represents the code change. 
Words, extracted from the first line of log message, can be viewed as semantic labels provided
by developers. 
Specifically, we define a learning task to construct a prediction function
$\textbf{f}: P \rightarrow \mathcal{Y}$, where $y_i \in \mathcal{Y}$
indicates the set of words extracted 
from the first line of the log message of the patch $p_i \in P$. 
The prediction function $\textbf{f}$ is learned by minimizing the differences between the
predicted and actual words chosen to describe the patch. 
After the prediction function $\textbf{f}$ is learned, for each patch, we can obtain its code change
vector from the intermediate output between the feature extraction and
feature fusion layers (see Figure~\ref{fig:framework}). 
We explain the
details of each part in the following subsections.


\subsection{Preprocessing}
\label{sec:preprocessing}
The code change of the given patch includes changes made to one or more files. Each changed file contains a set of lines of removed code and added code. We process the code change of each patch by the following steps:
\begin{itemize}
	\item \textbf{Split the code change based on the affected files.} We first separate the information about the code change to each changed file into a separate code document (i.e., $\text{File}_1$, $\text{File}_2$, etc., see Figure~\ref{fig:framework}).
	\item \textbf{Tokenize the removed code and added code lines.} For
          the changes affecting each changed file, we employ the NLTK library~\cite{bird2004nltk} for natural language processing (NLP) to parse its removed code lines or added code lines into a sequence of words. 
	We ignore blank lines in the changed file.  
	\item \textbf{Construct a code vocabulary.} Based on the code changes of the patches in the training data, we build a vocabulary $\mathcal{V}^{\text{C}}$. This vocabulary contains the set of code tokens that appear in the code changes of the collection of patches. 
\end{itemize}

At the end of this step, all the changed files of the given patch are extracted from the code changes and they are fed to the input layer of our framework for further processing.

\subsection{Input Layer}
\label{sec:input}
A code change may include changes to multiple files; the changes to
each file may contain changes to different hunks; and each hunk contains a list of removed and/or added code lines. To preserve this structural information, in each changed file, we represent the removed (added) code as a three-dimensional matrix, i.e., $\mathcal{B} \in \mathbb{R}^{\mathcal{H} \times \mathcal{L} \times W}$, where $\mathcal{H}$ is the number of hunks, $\mathcal{L}$ is the number of removed (added) code lines for each hunk, and $W$ is the number of words in each removed (added) code line in the affected file. We use $\mathcal{B}_r$ and $\mathcal{B}_a$ to denote the three-dimensional matrix of the removed and added code respectively.

Note that each patch may contain a different number of affected files
($\mathcal{F}$), each file may contain a different number of hunks
($\mathcal{H}$), each hunk may contain a different number of lines
($\mathcal{L}$), and each line may contain a different number of words
($W$). For parallelization~\cite{hoang2019deepjit, kim2014convolutional}, each input
instance is padded or truncated to the same $\mathcal{F}$, $\mathcal{H}$,
$\mathcal{L}$, and $W$.

\subsection{Feature Extraction Layers}
\label{sec:feature_extract}
\begin{figure}[t!]
	\centering 
	\includegraphics[width=\columnwidth]{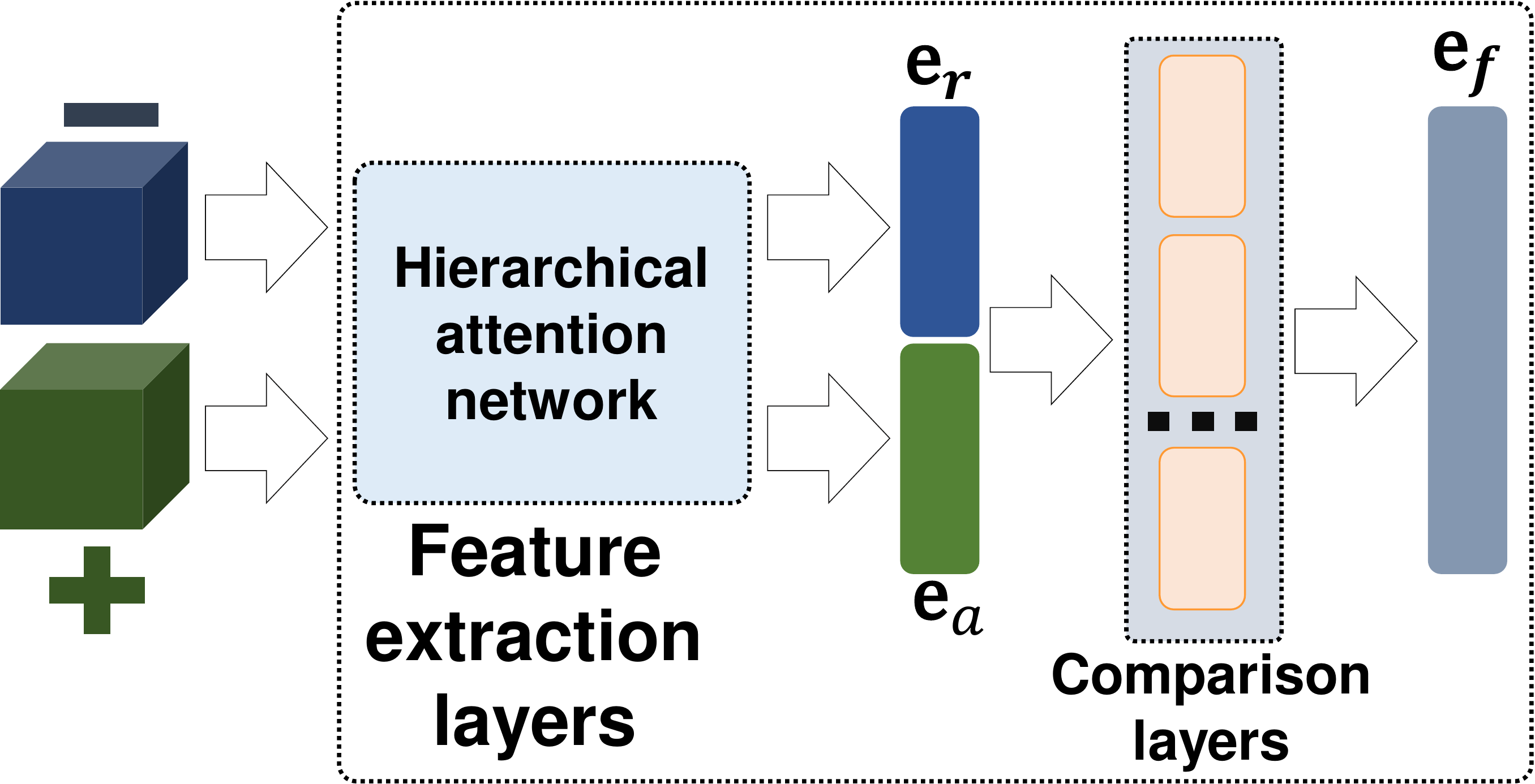}
	\caption{Architecture of the feature extraction layers for mapping the code change of the affected file in a given patch to an embedding vector. The input of the module is the removed code and added code of the affected file, denoted by ``-'' and ``+'', respectively.}
	\label{fig:feature_extraction}
\end{figure}

The feature extraction layers are used to automatically build an embedding vector representing the code change made to a given file in the patch. The embedding vectors of code changes to multiple files are then concatenated into a single vector representing the code change made by the patch. 

As shown in Figure~\ref{fig:feature_extraction}, for each affected file, the feature extraction layers take as input two matrices (denoted by ``-'' and ``+'' 
in Figure~\ref{fig:feature_extraction}) representing the removed code and added code, respectively. These two matrices are passed to the \textit{hierarchical attention network} 
to construct corresponding embedding vectors: $\textbf{e}_r$ representing the removed code and $\textbf{e}_a$ representing the added code (see Figure~\ref{fig:feature_extraction}). 
These two embedding vectors are fed to the \textit{comparison layers} to produce the vectors representing the difference between the removed code and the added code. 
These vectors are then concatenated to represent the code changes in each affected file. We present the hierarchical attention network and the comparison layers in the following sections.

\subsubsection{Hierarchical Attention Network}
\label{sec:HAN}
\begin{figure}[t!]
	\centering 
	\includegraphics[width=1.05\columnwidth]{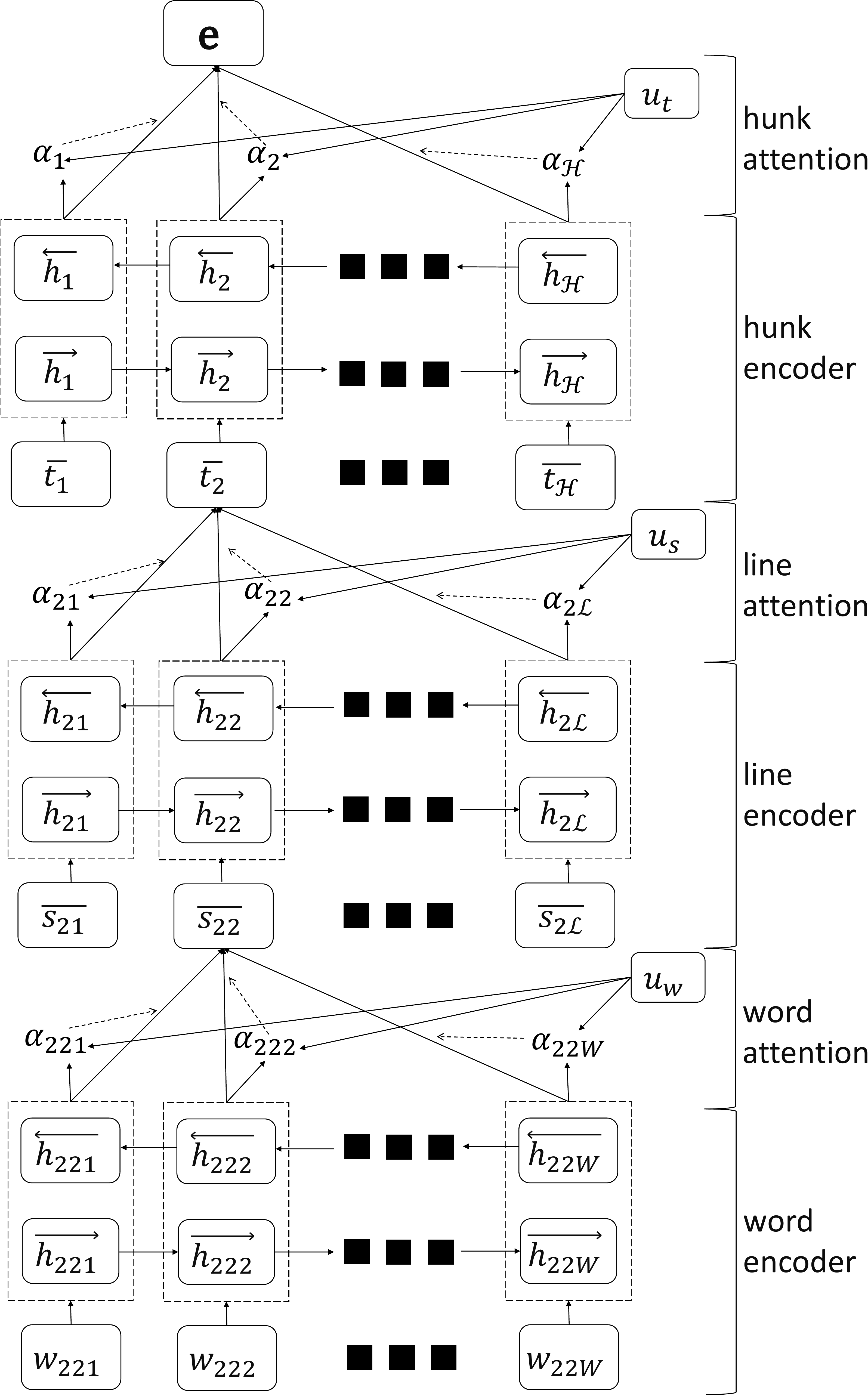}
	\caption{The overall framework of our hierarchical attention network (HAN). The HAN takes as input the removed (added) code of the affected file of a given patch and outputs the embedding vector (denoted by $\textbf{e}$) of the removed (added) code.}
	\label{fig:han}
\end{figure}

The architecture of our hierarchical attention network (HAN) is shown in Figure~\ref{fig:han}. 
A HAN takes the removed (added) code of an affected file of a given patch as an input and outputs the embedding vector representing the removed (added) code. 
Our HAN consists of several parts: a word sequence encoder, a word-level attention layer, a line encoder, a
line-level attention layer, a hunk sequence encoder, and a hunk attention layer. 

Suppose that the removed (added) code of the affected file contains a sequence of hunks  $\textbf{H} = [t_1, t_2, \dots, t_{\mathcal{H}}]$, each hunk $t_i$ includes a sequence of lines $[s_{i1}, s_{i2}, \dots, s_{i\mathcal{L}}]$, and each line $s_{ij}$ contains a sequence of words $[w_{ij1}, w_{ij2}, \dots, w_{ijW}]$. $w_{ijk}$ with $k \in [1, W]$ represents the word in the $j-$th line in the $i-$th hunk. 
Now, we describe how the embedding vector of the removed (added) code is built using the hierarchical structure. 

\noindent\textbf{Word encoder.} Given a line $s_{ij}$ with a sequence of words $w_{ijk}$ and a word embedding matrix $\textbf{W} \in \mathbb{R}^{|\mathcal{V}^C| \times d}$, where $\mathcal{V}^C$ is the vocabulary containing all words extracted from the code changes and $d$ is the dimension of the representation of word, we first build the matrix representation 
of each word in the sequence as follows:

\begin{equation} 
\overline{w_{ijk}} = \textbf{W}[w_{ijk}]
\end{equation}
where $\overline{w_{ijk}} \in \mathbb{R}^d$ indicates the vector representation of word $w_{ijk}$ in the word embedding matrix $\textbf{W}$. 
We employ a bidirectional 
GRU to summarize information from the context of a word in both directions~\cite{bahdanau2014neural}. 
To capture this contextual information,  
the bidirectional GRU includes a forward GRU that reads the line $s_{ij}$ from $w_{ij1}$ to $w_{ijW}$ and a backward GRU that reads the line $s_{ij}$ from $w_{ijW}$ to $w_{ij1}$. 

\begin{equation} 
\begin{split}
&\overrightarrow{h_{ijk}} = \overrightarrow{GRU}(\overline{w_{ijk}}), k \in [1, W] \\
&\overleftarrow{h_{ijk}} = \overleftarrow{GRU}(\overline{w_{ijk}}), k \in [W, 1]
\end{split}
\end{equation}

We obtain an annotation of a given word $w_{ijk}$ by concatenating the forward hidden state $\overrightarrow{h_{ijk}}$ and the backward hidden state $\overleftarrow{h_{ijk}}$ of this word, i.e., $h_{ijk} = [\overrightarrow{h_{ijk}} \oplus \overleftarrow{h_{ijk}}]$ ($\oplus$ is the concatenation operator). 
$h_{ijk}$ summarizes the word $w_{ijk}$ considering its neighboring words.


\noindent\textbf{Word attention.} Based on the intuition that not all words contribute equally to extract the ``meaning'' of the line, 
we use the attention mechanism to highlight
words important for predicting the content of the log message. 
The attention mechanism was previously used in source code summarization and was shown to be effective for encoding source code sequences~\cite{jiang2017automatically,leclair2019neural}. 
We also use the attention mechanism to form an embedding vector of the line. 
We first feed an annotation of a given word $w_{ijk}$ (i.e., $h_{ijk}$) through a fully connnected layer (i.e., $\mathcal{W}_w$) to get a hidden representation (i.e., $u_{ijk}$) of $h_{ijk}$ as follows: 
\begin{equation}
u_{ijk} = \text{ReLU}(\mathcal{W}_w h_{ijk} + b_w)
\end{equation}
where ReLU is the rectified linear unit activation function~\cite{nair2010rectified}, as it generally provides better performance in various deep learning tasks~\cite{dahl2013improving, anastassiou2011univariate}. 
Similar to Yang et al.~\cite{yang2016hierarchical}, we define a word context vector ($u_w$) that can be seen as a high level representation of the answer to the fixed query ``what is the most informative word'' over the words. 
The word context vector $u_w$ is randomly initialized and learned during the training process. 
We then measure the importance of the word as the similarity of $u_{ijk}$ with the word context vector $u_w$ and get a normalized importance weight $\alpha_{ijk}$ through a softmax function~\cite{bouchard2007efficient}:
\begin{equation}
    \alpha_{ijk} = \frac{\text{exp}(u_{ijk}^{\text{T}}u_w)}{\sum_k\text{exp}(u_{ijk}^{\text{T}}u_w)}
\end{equation}

For each line $s_{ij}$, its vector is computed as a weighted sum of the embedding vectors of the words based on their importance as follows:
\begin{equation}
    \overline{s_{ij}} = \sum_{k} \alpha_{ijk}h_{ijk} 
\end{equation}

\noindent\textbf{Line encoder.} Given a line vector (i.e., $\overline{s_{ij}}$), we also use a bidirectional GRU to encode the line as follows: 

\begin{equation} 
\begin{split}
&\overrightarrow{h_{ij}} = \overrightarrow{GRU}(\overline{s_{ij}}), j \in [1, \mathcal{L}] \\
&\overleftarrow{h_{ij}} = \overleftarrow{GRU}(\overline{s_{ij}}), j \in [\mathcal{L}, 1]
\end{split}
\end{equation}

Similar to the word encoder, we obtain an annotation of the line $s_{ij}$ by concatenating the forward hidden state $\overrightarrow{h_{ij}}$ and backward hidden state $\overleftarrow{h_{ij}}$ of this line. 
The annotation of the line $s_{ij}$ is denoted as  $h_{ij} = [\overrightarrow{h_{ij}} \oplus \overleftarrow{h_{ij}}]$, which summarizes the line $s_{ij}$ considering its neighboring lines.

\noindent\textbf{Line attention.} We use an attention mechanism to learn the important lines to be used to form a hunk vector as follows:
\begin{equation}
u_{ij} = \text{ReLU}(\mathcal{W}_s h_{ij} + b_s)
\end{equation}

\begin{equation}
    \alpha_{ij} = \frac{\text{exp}(u_{ij}^{\text{T}}u_s)}{\sum_j\text{exp}(u_{ij}^{\text{T}}u_s)}
\end{equation}

\begin{equation}
    \overline{t_{i}} = \sum_{j} \alpha_{ij}h_{ij} 
\end{equation}
$\mathcal{W}_s$ is the fully connected layer to which we need to feed an annotation of the given line (i.e., $s_{ij}$). 
We define $u_s$ as the line context vector that can be seen as a high level representation of the answer to the fixed query ``what is the informative line'' over the lines. $u_s$ is randomly initialized and learned during the training process. $\overline{t_i}$ is the hunk vector of the $i$-th hunk in the removed (added) code. 

\noindent\textbf{Hunk encoder.} Given a hunk vector $\overline{t_i}$, we again use a bidirectional GRU to encode the hunk as follows: 
\begin{equation}
    \begin{split}
    &\overrightarrow{h_{i}} = \overrightarrow{GRU}(\overline{t_{i}}), t \in [1, \mathcal{H}] \\
    &\overleftarrow{h_{i}} = \overleftarrow{GRU}(\overline{t_{i}}), t \in [\mathcal{H}, 1]
\end{split}
\end{equation}

An annotation of the hunk $t_i$ is then obtained by concatenating the forward hidden state $\overrightarrow{h_{i}}$ and the backward hidden state $\overleftarrow{h_{i}}$, i.e., $h_i = [\overrightarrow{h_{i}}, \overleftarrow{h_{i}}]$. $h_{i}$ summarizes the hunk $t_{i}$ considering the other hunks around it. 

\noindent\textbf{Hunk attention.} We again use an attention mechanism to learn important hunks used to form an embedding vector of the removed (added) code as follows:

\begin{equation}
u_{i} = \text{ReLU}(\mathcal{W}_{\text{h}} h_{i} + b_{\text{h}})
\end{equation}

\begin{equation}
    \alpha_{i} = \frac{\text{exp}(u_{i}^{\text{T}}u_t)}{\sum_i\text{exp}(u_{i}^{\text{T}}u_t)}
\end{equation}

\begin{equation}
    \textbf{e} = \sum_{i} \alpha_{i}h_{i} 
\end{equation}
$\mathcal{W}_{\text{h}}$ is the fully connected layer used to feed an annotation of a given hunk (i.e., $h_i$). $u_t$ is the hunk context vector that can be seen as a high level representation of the answer to the fixed query ``what is the informative hunk'' over the hunks. 
Similar to $u_w$ and $u_s$, $u_t$ is randomly initialized and learned during the training process. \textbf{e}, collected at the end of this part, is the embedding vector of the removed (added) code. 
For convenience, we denote $\textbf{e}_r$ and $\textbf{e}_a$ as the embedding vectors of the removed code and added code, respectively. 

\subsubsection{Comparison Layers}
\label{sec:comparison_layer}
\begin{figure}[t!]
	\centering 
	\includegraphics[width=\columnwidth]{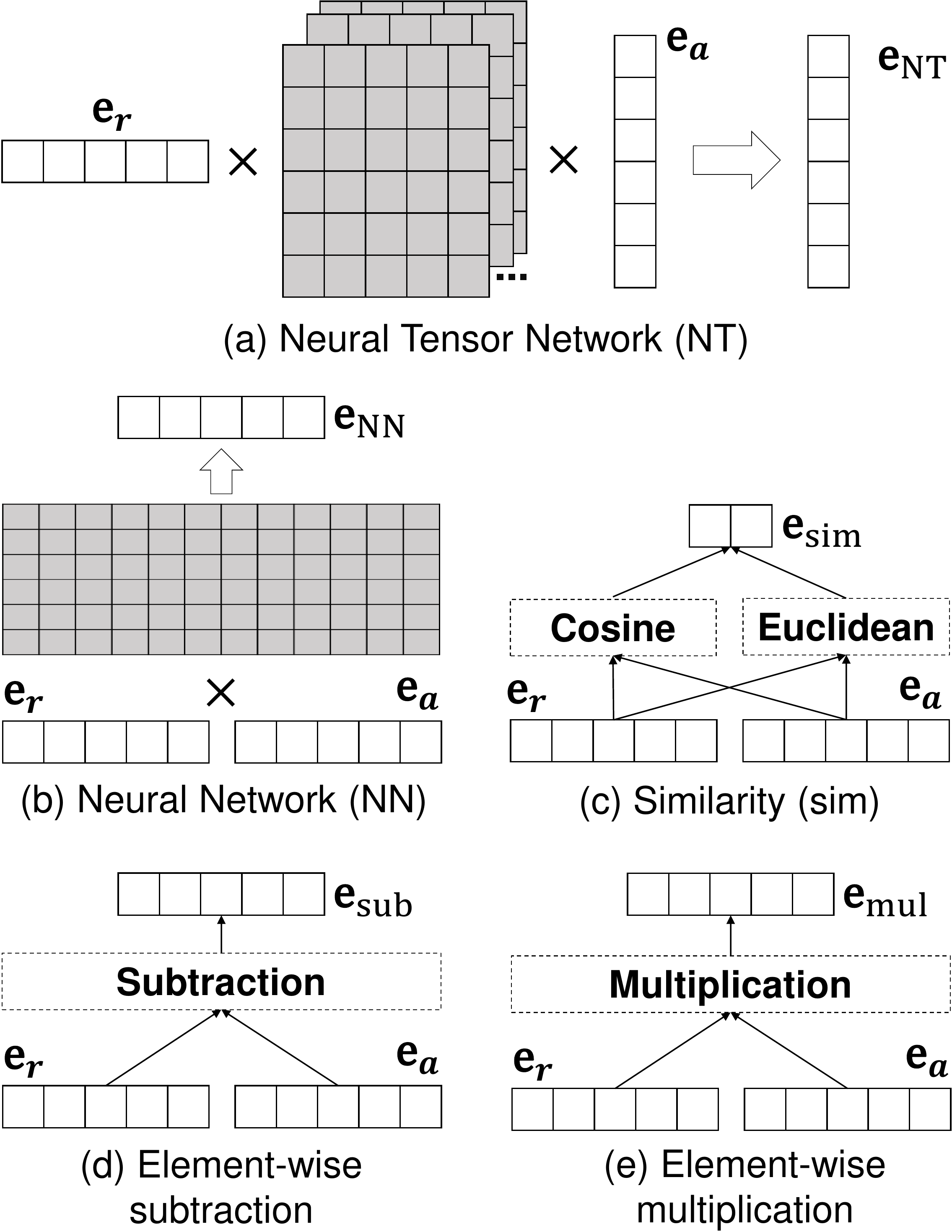}
	\caption{A list of comparison functions in the comparison layers.}
	\label{fig:comparison}
\end{figure}

The goal of the comparison layers is to build the vectors that capture the differences between the removed code and added code of the affected file in a given patch. 
We use multiple comparison functions~\cite{wang2017compare} to represent different angles of comparison. 
These comparison functions were previously used in a question answering
task.
The comparison layers take as input the embedding vectors of the removed code and added code (denoted by $\textbf{e}_r$ and $\textbf{e}_a$, respectively) and output the vectors representing the difference between the removed code and the added code. 
These vectors are then concatenated to represent an embedding vector of the affected file in the given patch.  Figure~\ref{fig:comparison} shows the five comparison functions used in the comparison layers to capture the difference between the removed code and added code. 
We briefly explain these comparison functions in the following paragraphs. 

\noindent\textbf{(a) Neural Tensor Network.} Inspired by previous works in 
visual question answering~\cite{bai2018deep}, 
we employ a neural tensor network~\cite{socher2013recursive} as follows: 

\begin{equation}
    \textbf{e}_{\text{NT}} = \text{ReLU}(\textbf{e}_r^{\text{T}} \textbf{T}^{[1, \dots, n]} \textbf{e}_a + b_{\text{NT}})
\end{equation}
$\textbf{T}^i \in \mathbb{R}^{n \times n}$ is a tensor and $b_{\text{NT}}$ is the bias value. These parameters are learned during the training process. Note that both the removed code and added code have the same dimension (i.e., $\textbf{e}_r \in \mathbb{R}^{n}$, $\textbf{e}_a$). 

\noindent\textbf{(b) Neural Network.} We consider a simple feed forward neural network~\cite{svozil1997introduction}. The output is computed as follows: 

\begin{equation}
    \textbf{e}_{\text{NN}} = \text{ReLU}(\textbf{W}[\textbf{e}_a \oplus \textbf{e}_r] + b_{\text{NN}})
\end{equation}
$\oplus$ is the concatenation operator, the matrix $\textbf{W} \in \mathbb{R}^{n \times 2n}$, and the bias value $b_{\text{NN}}$ are parameters to be learned. 

\noindent\textbf{(c) Similarity.} We employ two different similarity measures, euclidean distance and cosine similarity, to capture the similarity between the removed code and added code as follows:
\begin{equation}
\begin{split}
\textbf{e}_{\text{sim}} &=   \text{EUC}(\textbf{e}_r, \textbf{e}_a) \oplus \text{COS}(\textbf{e}_r, \textbf{e}_a)\\
\text{EUC}(\textbf{e}_r, \textbf{e}_a) &= ||\textbf{e}_r - \textbf{e}_a||_2 \\
\text{COS}(\textbf{e}_r, \textbf{e}_a) &= \frac{\textbf{e}_r \textbf{e}_a}{||\textbf{e}_r|| ||\textbf{e}_a||} \\
\end{split}
\end{equation}
$\text{EUC}(\cdot)$ and $\text{COS}(\cdot)$ are the euclidean distance and cosine similarity, respectively. Note that $\textbf{e}_{\text{sim}}$ is a two-dimensional vector. 

\noindent\textbf{(d) Element-wise subtraction.} We simply perform a subtraction between the embedding vector of the removed code and the embedding vector of the added code. 
\begin{equation}
	\textbf{e}_{\text{sub}} = \textbf{e}_r - \textbf{e}_a
\end{equation}

\noindent\textbf{(e) Element-wise multiplication.} We perform element-wise multiplication for the embedding vectors of the removed code and added code.
\begin{equation}
\textbf{e}_{\text{mul}} = \textbf{e}_r \odot \textbf{e}_a
\end{equation}
where $\odot$ is the element-wise multiplication operator.

The vectors resulting from applying these five different comparison functions are then concatenated to represent the embedding vector of the affected file (denoted by $\textbf{e}_{\texttt{f}_\text{i}}$) in the given patch as follows: 
\begin{equation}
	\textbf{e}_{\texttt{f}_\text{i}} = \textbf{e}_\text{NT} \oplus \textbf{e}_\text{NN} \oplus \textbf{e}_\text{sim} \oplus \textbf{e}_\text{sub} \oplus \textbf{e}_\text{mul}
\end{equation} 
where ${\texttt{f}_\text{i}}$ is the i-th file of the code change in the given patch. 

\subsection{Feature Fusion and Word Prediction Layers}
\label{sec:output_layer}

\begin{figure}[t!]
	\center
	\includegraphics[width=\columnwidth]{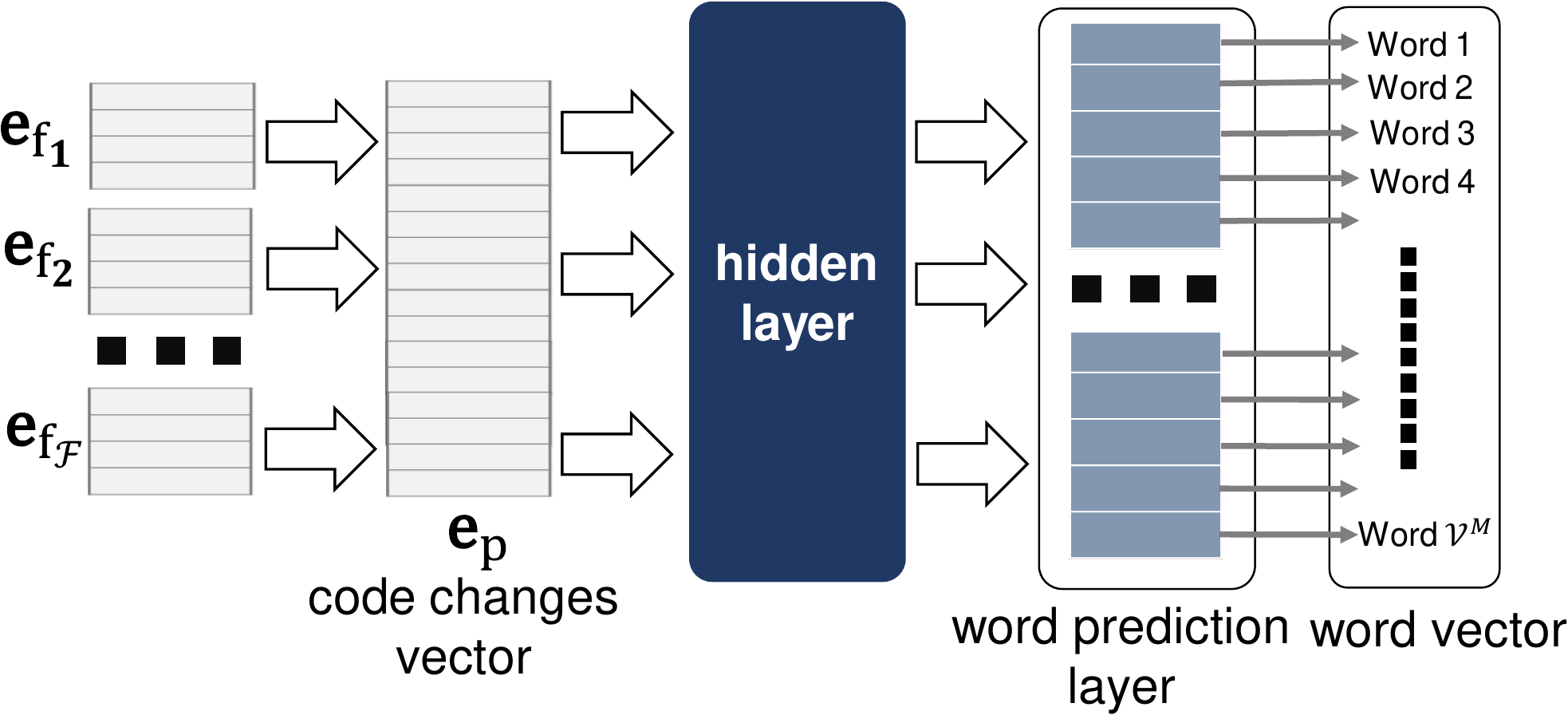}
	\caption{\textcolor{black}{The details of the red dashed box in Figure~\ref{fig:framework}. It takes as input a list of embedding vectors of the affected files of a given patch (i.e., $\textbf{e}_{\texttt{f}_1}$, $\textbf{e}_{\texttt{f}_2}$, $\dots$, $\textbf{e}_{\texttt{f}_\mathcal{F}}$). $\textbf{e}_\text{p}$ is the vector representation of the code change and is fed to a hidden layer to produce the word vector (i.e., the probability distribution over words). $\mathcal{V}^{M}$ is a set of words extracted from the first line of the log messages.}} 
	\label{fig:patch_vector}
\end{figure}

Figure~\ref{fig:patch_vector} shows the details of the part of the architecture shown inside the red (dashed) box in Figure~\ref{fig:framework}. The inputs of this part are the list of embedding vectors (i.e., $\textbf{e}_{\texttt{f}_1}$, $\textbf{e}_{\texttt{f}_2}$, $\dots$, $\textbf{e}_{\texttt{f}_\mathcal{F}}$) representing the features extracted from the list of affected files of a given patch. These embedding vectors are concatenated to construct a new embedding vector ($\textbf{e}_\text{p}$) representing the code change in a given patch as follows:

\begin{equation}
    \textbf{e}_\text{p} = \textbf{e}_{\texttt{f}_1} \oplus \textbf{e}_{\texttt{f}_2} \oplus \dots \oplus \textbf{e}_{\texttt{f}_\mathcal{F}}
\end{equation}

We pass the embedding vector ($\textbf{e}_\text{p}$) into a hidden layer (a fully connected layer) to produce a vector $\textbf{h}$:
\begin{equation} 
\label{eq:fully_layer}
\textbf{h} = \alpha(\textbf{w}_\textbf{h}\textbf{e}_\text{p} + b_\textbf{h})
\end{equation}
where $\textbf{w}_\textbf{h}$ is the weight matrix used to connect the embedding vector $\textbf{e}_\text{p}$ with the hidden layer and $b_\textbf{h}$ is the bias value. Finally, the vector $\textbf{h}$ is passed to a word prediction layer to produce the following:

\begin{equation} 
\label{eq:score_output}
\textbf{o} = -\textbf{h} \textbf{w}_\textbf{o} 
\end{equation}
where $\textbf{w}_\textbf{o}$ is the weight matrix between the hidden layer and the word prediction layer, and $\textbf{o} \in \mathbb{R}^{|\mathcal{V}^{M}| \times 1}$ ($\mathcal{V}^{M}$ is a set of words extracted from the first line of log messages). We then apply the sigmoid function~\cite{bouchard2007efficient} to get the probability distribution over words as follows:

\begin{equation}
\label{eq:prob_output}
\textbf{p}(o_i|p_i) =  \frac{1}{1 + \exp(o_i)}
\end{equation}
where $o_i \in \textbf{o}$ is the probability score of the $i^{th}$ word and $p_i$ is the patch that we want to assign words to.

\subsection{Parameter Learning}
\label{sec:parameters}

Our model involves the following parameters: the word embedding matrix of code changes, the hidden states in the different encoders (i.e., the word encoder, line encoder, and hunk encoder), the context vectors of words, lines, and hunks, the weight matrices and the bias values of the neural tensor network and the neural net in the comparison layers, and the weight matrices and the bias values of the hidden layer and the word prediction layer. After these parameters are learned, the vector representation of the code change of each patch can be determined. These parameters are learned by minimizing the following objective function:
\begin{equation} 
\begin{split}
\mathcal{O} &= \sum_{y_i \in \textbf{y}} ( y_i \times  -\log(\textbf{p}(o_i|p_i))  + (1 - y_i) \\
& \times -\log(1 - \textbf{p}(o_i|p_i)) ) + \frac{\lambda}{2} \|\theta\|_{2}^{2}
\end{split}
\label{eq:cost}
\end{equation}
where $\textbf{p}(o_i|p_i)$ is the predicted word probability defined in Equation~\ref{eq:prob_output}, 
$y_i = \{0, 1\}$ indicates whether the $i$-th word is part of the log message of the patch 
$p_i$, and $\theta$ are all parameters of our model. 
The regularization term, $\frac{\lambda}{2} \|\theta\|_{2}^{2}$, is used to prevent overfitting in the training process~\cite{caruana2001overfitting}. 
We employ the dropout technique~\cite{srivastava2014dropout} to improve the robustness of \PatchVec. 
Since Adam~\cite{kingma2014adam} has been shown to be computationally efficient and require low memory consumption, we use it to minimize the objective function (i.e., Equation~\ref{eq:cost}).
We also use backpropagation~\cite{hagan1994training}, a simple implementation of the chain rule of partial derivatives, to efficiently update the parameters during the training process.

%% file: experiment.tex
\section{Experiments}
\label{sec:experiment}

The goal of this work is to build a representation of code changes that can be applied to multiple tasks.
To evaluate the effectiveness of this representation, we employ our framework, namely \PatchVec, on three different tasks, i.e., 
log message generation~\cite{liu2018neural}, bug fixing patch identification~\cite{hoang2019patchnet} and just-in-time defect prediction~\cite{hoang2019deepjit}. 

In the first task of log message generation, we use the vector
representation of code changes, extracted by \PatchVec, to find a patch
that is most similar to another. 
For the other two tasks, \PatchVec~ is used to extract additional features that are input to the models of bug fixing patch identification and just-in-time defect prediction. We compare the resulting performance with and without using our code change vector. We next elaborate on the three tasks, the baselines, and results. 

\subsection{Task 1: Log Message Generation}
\subsubsection{Problem Formulation}

While we learn representations of code changes with the aid of log messages, we also study the task of generating log messages from code changes. 
Developers do not always write high-quality log messages. 
Dyer et al.~\cite{dyer2013boa} reported that around 14\% of log messages in 23,000 Java projects on SourceForge\footnote{https://sourceforge.net/} were empty. Log messages are important for program comprehension and understanding the evolution of software, 
therefore this 
motivates the need for the automatic generation of log messages.
In this task, given the code change of a given patch, we aim to produce a brief log message summarizing it. 

\subsubsection{State-of-the-art Approach}

The state-of-the-art approach is \textit{NNGen}~\cite{liu2018neural}, which takes as input a new code change with an unknown log message and a training dataset (patches), and outputs a log message for the new code change. NNGen first extracts code changes from the training set. 
Each code change in the training set and the new code change are represented as vectors in the form of a ``bag-of-words''~\cite{manning2010introduction}. NNGen then calculates the cosine similarity between the vector of the new code change and the vector of each code change in the training set, and selects the top-k nearest neighbouring code changes in the training dataset. From these k nearest neighbours, the BLEU-4 score~\cite{papineni2002bleu} is computed between each of the code changes in the top-k and the new code change with an unknown log message. A log message of the code change in the top-k with the highest BLEU-4 score is reused as the log message of the new code change. 

The BLEU-4 score is a measure used to evaluate the quality of machine translation systems, measuring the closeness of a translation to a human translation. It is computed as follows:

\begin{center}
$B L E U=B P \cdot \exp \left(\sum_{n=1}^{N} \frac{1}{N} \log \left(p_{n}\right)\right)$

$B P=\left\{\begin{array}{cl}{1} & {\text { if } c>r} \\ {e^{(1-r / c)}} & {\text { if } c \leq r}\end{array}\right.$
\end{center}

$N$ is the maximum number of N-grams. Following the previous work~\cite{liu2018neural}, we select $N = 4$.
$p_{n}$ is the ratio of length $n$ subsequences that are present in both  the output and reference translation. 
$ B P $ is a brevity penalty to penalize short output sentences.
Finally, $c$ is the length of the output translation and $r$ is the length of the reference translation.

A deep learning approach was previously proposed for this task by Jiang et al.~\cite{jiang2017automatically}, however, it underperformed the simpler baseline NNGen. In this study, we refer to their work as \textit{NMT}. 
Their approach modelled this task as a neural machine translation task, translating the code change to a target log message. Like our work, they proposed an attention-based model, however, our work differs from theirs as ours incorporates the structure of code changes. Liu et al.~\cite{liu2018neural} investigated the performance of Jiang et al.'s attention model; they found that once they remove trivial and automatically-generated messages, the performance of the model decreased significantly, suggesting that this model does not generalize. 

\subsubsection{Our Approach}

To use CC2Vec for this task, we propose \textit{LogGen}. Similar to the nearest neighbours approach used by Liu et al.~\cite{liu2018neural}, LogGen reuses and outputs a log message from the training set. 
However, instead of treating each code change as a bag of words, LogGen uses code change vectors produced by CC2Vec. CC2Vec is first trained over the training dataset. 
Given a new code change from the test dataset with an unknown log message, we find the code changes with a known log message that have the closest CC2Vec vector. 
Like Liu et al.~\cite{liu2018neural}, after identifying the closest code changes, we reuse the log message as the output. 

\subsubsection{Experimental Setting}

The purpose of evaluating CC2Vec on this task is to determine if the code change representations received from CC2Vec outperform the naive representation used by Liu et al.~\cite{liu2018neural}.
Jiang et al.~\cite{jiang2017automatically} originally collected and filtered the commits to construct the original dataset.  
Another version of the dataset was used by Liu et al.~\cite{liu2018neural}, who modified the original dataset. 

Jiang et al. extracted a total of 2 million patches
from the 1K most starred Java projects.
They collected the first line of each log message.
To normalize the dataset, patch ids and issue ids were removed from the code changes and log messages. 
Patches were filtered to remove merges, rollbacks, and patches that were too long.
The log messages that do not conform
to verb-direct-object pattern, e.g. ``delete a method'', are also removed.
After filtering, the dataset contains 32K patches.

Still, even with all this cleaning, Liu et al.~\cite{liu2018neural} investigated the dataset and found that there were many patches with bot messages and trivial messages. Bot messages refer to messages produced automatically by other development tools, such as continuous integration bots. 
Trivial messages refer to messages containing only information that can be obtained by looking at the names of the changed files (e.g. ``modify dockerfile''). Such messages are of low quality and Liu et al.\ used regular expressions to locate and remove these patches.

We used the original dataset of Jiang et al.~\cite{jiang2017automatically} and the cleaned dataset of Liu et al.~\cite{liu2018neural} for evaluation.
While the original dataset consists of a training dataset of 30K patches and a testing dataset of 3K patches, the cleaned dataset consists of a training dataset of 22K patches and a testing dataset of 2.5K patches.
To compare the different approaches, 
we use BLEU-4 to evaluate each approach since this was used in both previous works.

\subsubsection{Results}

\begin{table}
    \caption{Performance of each approach on the original and cleaned dataset reported in BLEU-4}
      \label{tab:commit_log_gen}
      \begin{tabular}{lccc}
        \hline
        & LogGen & NNGen & NMT \\ 
        \hline
        \textit{Original} & \textbf{43.20} & 38.55 & 31.92 \\
        \hline
        \textit{Clean} & \textbf{20.48} & 16.42 & 14.19 \\
        \hline
    \end{tabular}
  \end{table}

We report the performance of LogGen, NNGen and NMT in Table \ref{tab:commit_log_gen}. 
LogGen outperforms both NNGen and NMT. 
The \textit{Clean} dataset refers to the dataset which Liu et al. filtered out patches with bot and trivial log messages.
On this dataset, LogGen outperforms NNGen and NMT by a BLEU-4 score of 4.06 and 6.29 respectively.
LogGen \textit{improves} over the performance of NNGen by 24.75\%, a greater improvement than NNGen's improvement over NMT of 15.70\% . On the original dataset collected by Jiang et al., LogGen outperforms NNGen and NMT by a BLEU-4 score of 4.65 and 11.28. 
These results indicate that LogGen can improve over the performance of NNGen and NMT by 12.06\% and 2.07\% in terms of the BLEU-4 score respectively.

Thus, we conclude that the log messages retrieved by LogGen are closer in quality to a human translation than those retrieved by NNGen and the log messages generated by NMT. This suggests that CC2Vec produces vector representations of patches that correlate to the meaning of the patch more strongly than a bag-of-words.

\subsection{Task 2: Bug Fixing Patch Identification}
\label{sec:experiment_linux_bugfix}

\subsubsection{Problem Formulation}

Software requires continuous evolution to keep up with new requirements, but this also introduces new bugs.
Backporting bugfixes to older versions of a project may be required when a legacy code base is supported. 
For example, Linux kernel developers regularly backport bugfixes from the latest version to older versions that are still under support. However, the maintainers of older versions may overlook relevant patches in the latest version. 
Thus, an automated method to identify bug fixing patches may be helpful.  
We treat the problem as a binary classification problem, in which each patch is labelled as a bug-fixing patch or not. Given the code change and log message, we produce one of the two labels as the output.

\subsubsection{State-of-the-art Approach}

The state-of-the-art approach is \textit{PatchNet}~\cite{hoang2019patchnet}, which represents the removed (added) code as a three dimensional matrix. The dimensions of the matrix are the number of hunks, the number of lines in each hunk, and the number of words in each line. PatchNet employs a 3D-CNN~\cite{ji20123d} that automatically extracts features from this matrix. 
Unfortunately, the 3D-CNN lacks a mechanism to identify important words, lines, and hunks. To address this limitation, we propose a specialized hierarchical attention neural network  to quantify the importance of words, lines, and hunks in our model (\PatchVec). 


Another approach was proposed by Tian et al.~\cite{tian2012identifying} that combines Learning from Positive and Unlabelled examples (LPU)~\cite{lee2003learning} and Support Vector Machine (SVM)~\cite{joachims1999svmlight} to build a patch classification model. Unlike CC2Vec, this approach requires the use of manually selected features. These features include word features, which is a ``bag-of-words'' extracted from log messages, and 52 features, manually extracted from the code change (e.g., the number of loops added in a patch and if certain words appear in the log message).

\subsubsection{Our Approach}

CC2Vec is first used to learn a distributed representation of code changes on the whole dataset. 
All patches from the training and test dataset are used since 
the log messages of the test dataset are not the target of the task. 
Next, we integrate these vector representations of the code changes with the two existing approaches. 
To use \PatchVec~ in PatchNet, we concatenate the vector representation of the code change extracted by \PatchVec~ with the two embedding vectors extracted from the log message and code change  by PatchNet  to form a new embedding vector. 
The new embedding vector is fed into PatchNet's classification module to predict whether a given patch is a bug fixing patch.
For the approach proposed by Tian et al.~\cite{tian2012identifying} which uses an SVM as the classifier, we pass the vectors produced by CC2Vec from the code change into the SVM as features. 

\subsubsection{Experimental Setting}

The goal of this task is to investigate if CC2Vec helps existing approaches to effectively classify bug-fixing patches. 
We use the dataset of Linux kernel bug-fixing patches used in the PatchNet paper. This dataset consists of 42K bug-fixing patches and 40K non-bug-fixing patches collected from the Linux kernel versions v3.0 to v4.12, released in July 2011 and July 2017 respectively. Patches in this dataset are limited to 100 lines of changed code, in line with the Linux kernel stable patch guidelines. 
The non-bug-fixing patches are selected such that they have a similar size, in terms of the number of files and the number of modified lines, as the bug-fixing patches. Following the PatchNet paper, we use 5-fold cross-validation for the evaluation.

To compare the performance of the approaches, we employ the following metrics: 
\begin{itemize}[leftmargin=*]
    \item \textit{Accuracy}: The ratio of correct predictions to the total number of predictions. 
    \item \textit{Precision}: The ratio of correct predictions of bug-fixing patches to the total number of bug-fixing patch predictions
    \item \textit{Recall}: The ratio of correct predictions of bug-fixing patches to the total number of bug-fixing patches. 
    \item \textit{F1}: Harmonic mean between precision and recall.
    \item \textit{AUC}: Area under the curve plotting the true positive rate against the false positive rate. AUC values range from 0 to 1, with a value of 1 indicating perfect discrimination.
\end{itemize}
These metrics were also used in previous studies on this task.

\subsubsection{Results}

\begin{table}
    \caption{Evaluation of  the approaches on the bug-fixing patch identification task}
      \label{tab:buggy_commit}
      \begin{tabular}{lccccc}
        \hline
        & Acc. & Prec. & Recall & F1 & AUC \\ 
        \hline 
        LPU-SVM & 73.1 & 75.1 & 71.6 & 73.3 & 73.1  \\ 
        \hline 
        LPU-SVM + CC2Vec &  \textbf{77.1} & \textbf{77.2} & \textbf{79.8} & \textbf{78.5} & \textbf{76.2}  \\
        \hline
        PatchNet & 86.2 & 83.9 & 90.1 & 87.1 & 86.0 \\
        \hline 
        PatchNet + CC2Vec & \textbf{90.7} & \textbf{91.6} & \textbf{90.1} & \textbf{90.9} & \textbf{91.6} \\
        \hline
    \end{tabular}
  \end{table}

We report the performance of the different approaches in Table \ref{tab:buggy_commit}. 
We observe that the best performing approach is PatchNet augmented with CC2Vec. 
For both Tian et al.'s model (LPU-SVM) and PatchNet, the versions augmented with CC2Vec outperform the original versions. 
Specifically, CC2Vec helps to improve the best performing baseline (i.e, PatchNet) by 5.22\%, 9.18\%, 4.37\%, and 6.51\% in terms of accuracy, precision, F1, and AUC. CC2Vec also helps to improve the performance of LPU-SVM by 5.47\%, 2.80\%, 11.45\%, 7.09\%, and 4.24\% in accuracy, precision, recall, F1, and AUC. This suggests that CC2Vec can learn patch representations that are general and useful beyond the task it was trained on.

\subsection{Task 3: Just-in-Time Defect Prediction}
\subsubsection{Problem Formulation}

The task of just-in-time (JIT) defect prediction refers to the identification of defective patches.
JIT defect prediction tools provide early feedback to software developers to optimize their effort for inspection, and have been used at large software companies~\cite{mockus2000predicting,shihab2012industrial,tantithamthavorn2015impact}. 
We model the task as a binary classification task, in which each patch is labelled as a patch containing a defect or not. 
Given a patch containing a code change and a log message with unknown label, we label the patch with one of the two labels.

\subsubsection{State-of-the-art Approach}
The state-of-the-art approach is \textit{DeepJIT}, proposed by Hoang et al.~\cite{hoang2019deepjit}. 
DeepJIT takes as input the log message and code change of a given patch and outputs a probability score to predict whether the patch is buggy. DeepJIT employs a Convolutional Neural Network (CNN)~\cite{kim2014convolutional} to automatically extract features from the code change and log message of the given patch. However, DeepJIT ignores information about the structure of the removed code or added code, instead relying on CNN to automatically extract such information.



\subsubsection{Our Approach}

Similar to the previous task (i.e., bug fixing patch identification), CC2Vec is first used to learn distributed representations of the code changes in the whole dataset. 
All patches from the training and test dataset are used since the log messages of the test dataset are not part of the predictions of the task. 
We then integrate CC2Vec with DeepJIT. 
To use CC2Vec with DeepJIT, for each patch, we concatenate the vector representation of the code change extracted by CC2Vec with two embedding vectors extracted from the log message and code change of the given patch extracted by DeepJIT to form a new embedding vector. 
The new embedding vector is fed into DeepJIT's feature combination layers, to predict whether the given patch is defective. 

\subsubsection{Experimental Setting}

The purpose of this task is to evaluate if CC2Vec can be used to augment existing approaches in effectively classifying defective patches. 
Our evaluation is performed on two datasets, the \textit{QT} and \textit{OPENSTACK} datasets, which contain patches collected from the QT and OPENSTACK software projects respectively by McIntosh and Kamei~\cite{mcintosh2017fix}. 
The QT dataset contains 25K patches over 2 years and 9 months while the OPENSTACK dataset contains 12K patches over 2 years and 3 months.
8\% and 13\% of the patches are defective in the QT dataset and the OPENSTACK datasets respectively. 
Like Hoang et al.~\cite{hoang2019deepjit}, we use 5-fold cross validation for the evaluation.


To compute the effectiveness of the approaches, we use the Area Under the receiver operator characteristics Curve (AUC), similar to the previous studies. 

\subsubsection{Results}

\begin{table}
    \caption{The AUC results of the various approaches}
      \label{tab:jit}
      \begin{tabular}{lcc}
        \hline
        & QT & OPENSTACK \\ 
        \hline
        DeepJIT & 76.8 & 75.1  \\
        \hline 
        DeepJIT + CC2Vec & \textbf{82.2} & \textbf{80.9}\\
        \hline 
    \end{tabular}
  \end{table}

The evaluation results for this task are reported in Table \ref{tab:jit}. 
The use of CC2Vec with DeepJIT improves the AUCS score of DeepJIT, from 76.8 and 75.1 to 82.2 and 80.9 on the QT and OPENSTACK datasets respectively. 
Specifically, CC2Vec helps to improve the AUC metric by 7.03\% and 7.72\% for the QT and OPENSTACK datasets, respectively, as compared to DeepJIT. 
This indicates that CC2Vec is effective in learning a useful representation of patches that an existing state-of-the-art technique can utilize. 

%% file: discussion.tex
\section{Discussion}
\label{sec:discussion}

\subsection{Ablation Study}
\label{sec:ablation}

\begin{table*}[t!]
  \centering
  \caption{Results of an ablation study} 
    \begin{tabular}{l|cc|cc|cc|cc}
    \hline
    & \multicolumn{2}{c|}{\textbf{Log generation (BLEU-4)}} & \multicolumn{2}{c|}{\textbf{Bug fix identification (F1)}} & \multicolumn{4}{c}{\textbf{Just-in-time defect prediction (AUC)}} \\
\cline{2-9}          & \textit{Clean} & Drops by (\%) & BFP & Drops by (\%) & QT & Drops by (\%) & OPENSTACK & Drops by (\%) \\
    \hline
    All$-$all & 18.30  & 10.64 & 87.1  & 4.18  & 77.4  & 5.84  & 76.7  & 5.19 \\
    All$-$NT & 19.36 & 5.47  & 88.7  & 2.42  & 79.8  & 2.92  & 79.2  & 2.10 \\
    All$-$NN & 19.80  & 3.32  & 88.8  & 2.31  & 80.1  & 2.55  & 79.5  & 1.73 \\
    All$-$sim & 20.41 & 0.34  & 90.2  & 0.77  & 81.9  & 0.36  & 80.5  & 0.49 \\
    All$-$sub & 20.13 & 1.71  & 89.6  & 1.43  & 80.7  & 1.82  & 80.1  & 0.99 \\
    All$-$mul & 20.25 & 1.12  & 89.7  & 1.32  & 81.1  & 1.34  & 80.5  & 0.49 \\
    \hline
    All   & \textbf{20.48} & 0     & \textbf{90.9} & 0     & \textbf{82.2} & 0     & \textbf{80.9} & 0 \\
    \hline
    \end{tabular}%
  \label{tab:ablation}%
\end{table*}%

Our approach involves five comparison functions for calculating the difference between the removed code and added code. 
To estimate the usefulness of comparison functions (see Section~\ref{sec:comparison_layer}), we conduct an ablation study on the three tasks: log message generation, bug fixing patch identification, and just-in-time defect prediction. Specifically, we first remove the comparison functions entirely and then remove these functions one-by-one. For each task, we compare the CC2Vec model and its six reduced variants: All$-$all (omit all comparison functions), All$-$NT (omit the neural network tensor comparison function), All$-$NN (omit the neural network comparison function), All$-$sim (omit the similarity comparison function), All$-$sub (omit the subtraction comparison function), and All$-$mul (omit the multiplication comparison function).


Table~\ref{tab:ablation} summarizes the results of our ablation test on three different tasks. We see that CC2Vec model always performs better than the reduced variants for all three tasks. This suggests that each comparison function plays an important role and omitting these comparison functions may greatly affect the overall performance. All$-$all (CC2Vec model without using any comparison functions) performs the worst. Among the five remaining variants (i.e., All$-$NT, All$-$NN, All$-$sim, All$-$sub, and All$-$mul), All-NT performs the worst. This suggests that the \textit{neural network tensor} comparison function is more important the other comparison functions (i.e., \textit{neural network}, \textit{similarity}, \textit{subtraction}, and \textit{multiplication}).

\subsection{Threats to Validity}
\label{sec:threats}
Threats to internal validity refer to errors in our experiments and experimenter bias. 
For each task, we reuse existing implementations of the baseline approaches whenever available.  
We have double checked our code and data, but errors may remain.

Threats to external validity concern the generalizability of our work.
In our experiments, we have studied only three tasks to evaluate the generality of CC2Vec. 
This may be a threat to external validity since CC2Vec may not generalize beyond the tasks that we have considered. 
However, each task involves different software projects and different programming languages. 
As such, we believe that there is minimal threat to external validity.
To minimize threats to construct validity, we have used the same evaluation metrics that were used in previous studies.

%% file: related_work.tex
\section{Related Work}
\label{sec:related_work}

There are many studies on the representation of source code, 
including recent studies proposing distributed representations for identifiers~\cite{efstathiou2019semantic}, APIs~\cite{nguyen2016mapping,nguyen2017exploring}, and software libraries~\cite{theeten2019import2vec}. A comprehensive survey of learning the representation of source code has been done by Allamanis et al.~\cite{allamanis2018survey}.

Some studies transform the source code into a different form, such as control-flow graphs~\cite{defreez2018path} and symbolic traces~\cite{henkel2018code}, or collect runtime execution traces~\cite{wang2017dynamic}, before learning distributed representations.
DeFreez et al.~\cite{defreez2018path} found function synonyms by learning embeddings through random walks of the interprocedural control-flow graph of a program. These embeddings are then used in a single downstream task of mining error-handling specifications. 
Henkel et al.~\cite{henkel2018code} described a toolchain to produce abstracted intraprocedural symbolic traces for learning word embeddings.
They experimented on a downstream task to find and repair bugs related to incorrect error codes. 
Wang et al.~\cite{wang2017dynamic} used execution traces to learn embeddings. They integrate their embeddings into a program repair system in order to produce fixes to correct student errors in programming assignments.
These studies differ from our work as we leverage natural language data as well as source code.

There have been other studies using deep learning of both source code and natural language data, for example, 
joint learning of embeddings for both text and source code to improve code search~\cite{gu2018deep}.
Other studies proposed approaches to learn distributed representations of source code on prediction tasks with natural language output.
Iyer et al.~\cite{iyer2016summarizing} proposed a model using LSTM networks with attention for code summarization, 
and Yin et al.~\cite{yin2018learning} trained a model to align source code to natural language text from StackOverflow posts. 
However, unlike our work, these studies do not use structural information of the source code. 

Several studies~\cite{hu2018deep,alon2019code2vec,alon2018codeseq,kovalenko2019pathminer} account for structural information but differ from our work. 
Hu et al.~\cite{hu2018deep} proposed an approach to use Sequence-to-Sequence Neural Machine Translation to generate method-level code comments. 
By prefixing the AST node type in each token and traversing the AST of methods such that the original AST can be unambiguously reconstructed, 
they convert the AST of each method into a sequence that preserves structural information. 
Alon et al. proposed code2vec~\cite{alon2019code2vec}, which represents code as paths in an AST, learning the vector representation of each AST path. 
They trained their model on the task of predicting a label, such as the method name, of the code snippet. 
In a later work, they proposed code2seq~\cite{alon2018codeseq}. 
Instead of predicting a single label, they generate a sequence of natural language words. 
Similar to our work, structural information of the input source code is encoded in the model's architecture, 
however, in these studies, 
the input code snippet is required to be parseable to build an AST. 

As our work focuses on the representation of software patches, 
we deliberately designed CC2Vec to not require parseable code in its input. 
This is done for two reasons. 
Firstly, a small but still significant proportion of patches may have compilation errors. 
A study by Beller et al. on Travis CI build failures revealed that about 4\% of Java project build failures are due to compilation errors~\cite{beller2017oops}. 
CC2Vec is designed to be usable even for these patches. 
Secondly, parsing will require the entire file with the changed code. Retrieving this information and parsing the entire file will be time consuming.

All the studies above proposed general representations of source code. 
The representations they learn, with the exception of DeFreez et al.~\cite{defreez2018path}, are of source code contained in a single function.
In contrast, we learn representations of code changes, which can contain modifications to multiple different functions, across multiple files. 

Several of the models related to code changes' representation were discussed in Section \ref{sec:experiment}. 
These models often do not model the hierarchical structure of a code change or require handcrafted features that may be specific to a single task~\cite{liu2018neural,jiang2017automatically,tian2012identifying,kamei2012large,aversano2007learning,kim2008classifying,kamei2016studying,mockus2000predicting,yang2015deep}.
%

Two techniques using deep neural networks, PatchNet~\cite{hoang2019patchnet} and DeepJIT~\cite{hoang2019deepjit}, are most similar to our work. 
However, as discussed earlier, our work differs from theirs in various ways.
A fundamental difference is in the generality of the techniques.
CC2Vec is not specific to a single task.
Rather, CC2Vec can be trained for multiple tasks, including both generative and classification tasks.
In fact, CC2Vec is orthogonal to these approaches. 
The objective of CC2Vec is to produce high quality representations of code changes that can be integrated into PatchNet, DeepJIT, and similar models. 
We showed in Section \ref{sec:experiment} that the performance of these models improves when they are augmented with the code change representation learned by CC2Vec. 

%% file: conclusion.tex
\section{Conclusion}
\label{sec:conclusion}

We propose CC2Vec, which produces distributed representations of code changes through a hierarchical attention network. 
In CC2Vec, we model the structural information of a code change and use the attention mechanism to identify important aspects 
of the code change with respect to the log message accompanying it. 
This allows CC2Vec to learn high-quality vector representations that can be used in existing state-of-the-art models on tasks involving code changes. 

We empirically evaluated CC2Vec on three tasks and demonstrated that approaches using or augmented with CC2Vec embeddings outperform existing state-of-the-art approaches that do not use the embeddings. Finally, we performed an ablation study to evaluate the usefulness of comparison functions. The results show that the comparison functions play an important role and omitting them in part or in full affects the overall performance. 

As future work, to reduce the threat to external validity, 
we will integrate of CC2Vec into other tools and experiments on other tasks involving software patches. 

\noindent \textbf{Package.} The replication package is available at \url{https://github.com/CC2Vec/CC2Vec}.

%% file: ack.tex
\noindent{\bf Acknowledgement.} This research was supported by the Singapore National Research Foundation (Award number: NRF2016-NRF-ANR003) and the ANR ITrans project.